\documentclass{article}

\usepackage{booktabs}
\usepackage{PRIMEarxiv}

\usepackage[utf8]{inputenc} 
\usepackage[T1]{fontenc}    
\usepackage{hyperref}       
\usepackage{url}            
\usepackage{booktabs}       
\usepackage{amsfonts}       
\usepackage{nicefrac}       
\usepackage{microtype}      
\usepackage{lipsum}
\usepackage{fancyhdr}       
\usepackage{graphicx}       
\graphicspath{{media/}}     
\usepackage{caption}
\usepackage{subcaption}
\usepackage{xcolor}
\newcommand{\eg}{\textit{e.g.}}

\newcommand{\tworow}[2]{\begin{tabular}{@{}l@{}}#1 \\ #2\end{tabular}}

\title{The Impact of Scanner Domain Shift on Deep Learning Performance in Medical Imaging: an Experimental Study
}

\author{%
Brian Guo$^{1,*}$, Darui Lu$^{2,*}$, Gregory Szumel$^{1}$, Rongze Gui$^{1}$, \\
\textbf{Tingyu Wang$^{1}$, Nicholas Konz$^{2,\dagger}$, Maciej A. Mazurowski$^{1,2,3,4,\dagger}$} \\
$^{1}$Dept. of Computer Science, 
$^{2}$Dept. of Electrical and Computer Engineering\\
$^{3}$Dept. of Radiology, 
$^{4}$Dept. of Biostatistics \& Bioinformatics, \\
Duke University, NC, USA \\
$^*$Equal contribution \quad $^\dagger$Advisory roles\\
\texttt{first.last@duke.edu} \\
}

\begin{document}
\maketitle

\begin{abstract}
\textbf{Purpose:}\\
Medical images acquired using different scanners and protocols can differ substantially in their appearance. This phenomenon, \textit{scanner domain shift}, can result in a drop in the performance of deep neural networks which are trained on data acquired by one scanner and tested on another. This significant practical issue is well-acknowledged, however, no systematic study of the issue is available across different modalities and diagnostic tasks.
\\\textbf{Materials and Methods:}\\
In this paper, we present a broad experimental study evaluating the impact of scanner domain shift on convolutional neural network performance for different automated diagnostic tasks. We evaluate this phenomenon in common radiological modalities, including X-ray, CT, and MRI. 
\\\textbf{Results:}\\
We find that network performance on data from a different scanner is almost always worse than on same-scanner data, and we quantify the degree of performance drop across different datasets. Notably, we find that this drop is most severe for MRI, moderate for X-ray, and quite small for CT, on average, which we attribute to the standardized nature of CT acquisition systems which is not present in MRI or X-ray. We also study how injecting varying amounts of target domain data into the training set, as well as adding noise to the training data, helps with generalization.
\\\textbf{Conclusion:}\\
Our results provide extensive experimental evidence and quantification of the extent of performance drop caused by scanner domain shift in deep learning across different modalities, with the goal of guiding the future development of robust deep learning models for medical image analysis.
\end{abstract}

\keywords{Deep Learning \and Medical Image Analysis \and Domain Shift \and Performance Survey}

\section{Introduction}
Medical image analysis has rapidly advanced through the use of deep learning \cite{chan2020deep,litjens2017survey,celard2023survey}, due to the ability to train highly flexible neural networks to use for various diagnostic tasks. However, due to their flexibility and data-driven nature, neural networks are susceptible to the problem of \textit{scanner domain shift} \cite{wang2022embracing}. 

When radiological images are acquired at different imaging sites, referred to as \textit{domains}, various attributes of the medical image acquisition pipeline such as scanner model and manufacturer, image acquisition parameters (\eg, echo and repetition times for MRI), and post-processing can differ between the two domains, resulting in images with different characteristics. Then, if a neural network is trained to perform a diagnostic task on the images from one site (\eg, cancer detection in breast MRI), 
and has learned to utilize certain image features for its predictions, these features may not present, to the same extent, in images taken in other domains, resulting in a drop in its performance compared to its in-domain diagnostic ability. Even if these image differences are barely visually perceptible, neural networks may still be highly sensitive to them \cite{szegedy2013intriguing,geirhos2018imagenet,konz2023reverse}.

In this paper, we perform a broad experimental study on the scanner domain shift problem in medical image analysis. In particular, we analyze the effect of 
changes in image characteristics due to scanner manufacturer on the performance of networks trained for different diagnostic tasks, on seven datasets from MRI, CT, and X-ray modalities. Understanding and addressing this variability is crucial for understanding the robustness of deep learning models in clinical settings to scanner domain shift, providing guidelines on the real-world applicability of these methods.

Our results show that in all but three of fourteen settings of a dataset, task, and source/target domains, scanner domain shift results in a loss of performance. Interestingly, we also see a general trend of domain shift severity with respect to imaging modality: on average, MRI tasks had the most severe domain shift issues, X-ray was moderate, and CT tasks were minimal, which we hypothesize is due to differing dependence of generated image features on the acquisition parameters that can change between scanners. Finally, we also evaluate the effects of gradually injecting target domain data into the training set or adding noise to training data on model generalization (Appendix \ref{app:additionalexp}). Our results show that scanner domain shift is an important, non-trivial factor that should not be ignored in deep learning-driven medical image analysis, especially for more vulnerable modalities.

In total, we present the first multi-modality, multi-body region systematic experimental study of the effect of scanner domain shift on neural network performance in medical imaging. Our results demonstrate that domain shift in medical imaging often has an effect on neural network performance---the severity of which differs depending on the imaging modality---and needs to be considered when training and evaluating networks for medical image analysis.

\section{Related Work}

Certain previous studies in the usage of deep learning for medical image analysis have focused on evaluating the performance of such techniques under various types of domain shift. For example, AlBadawy et al. \cite{albadawy2018deep} and Wang et al. \cite{wang2020inconsistent} demonstrated that neural networks trained for brain MRI tumor segmentation and mammogram classification, respectively, suffered from performance drops when given data from a different institution than that of the training set. In a further extension of this research, Yao et al. \cite{yao2020strong} conducted an extensive study using ten X-ray datasets to establish a baseline for domain adaptation and generalization, and observed consistent performance drops when trained models were given X-rays from unseen datasets. Additionally, Måartensson et al. \cite{maartensson2020reliability} found that MRI diagnostic models developed from homogeneous research cohorts often underperformed when applied to diverse clinical data, highlighting the challenges posed by out-of-distribution (OOD) data. A common theme across these studies is the significant decline in model performance in OOD scenarios, yet a notable limitation is that they each focus on single imaging modalities.

Building on these works, our study aims to understand the impact of domain shift specifically created by using differing scanners for image acquisition, across multiple imaging modalities. We note that as opposed to works such as \cite{guan2021domain}, this is not a survey of existing techniques for mitigating domain \textit{adaptation} (at the model, learned feature, or image level), but an experimental study of the effects of scanner-based domain shift on downstream task performance for models that have \textit{not} been adapted by such methods. As shown in the next section, we propose an experimental design that is as standardized as possible across all modalities and diagnostic tasks, with the goal of obtaining a general, universal understanding of scanner domain shift phenomena.

\section{Methods}
\subsection{Experimental Design and Models}
\label{sec:expdesign}

We gathered a total of seven datasets from the CT, MRI, and X-ray modalities to explore the impact of scanner domain shift on deep neural networks trained for various diagnostic tasks. Each dataset fulfills the following requirements:
    \begin{enumerate}
        \item The dataset must contain scanner information (i.e. the manufacturer of the scanner).
        \item The dataset needs enough data to sufficiently train and evaluate deep learning models.
        \item The dataset needs to be public for the sake of reproducibility.
    \end{enumerate}
All datasets and their respective diagnostic tasks will be individually introduced in the following section. For each dataset, we split the data into two scanner types (labeled 1 and 2), then for each scanner type into random subsets (by patient) of ${2}/{3}$ for training, ${1}/{6}$  for validation, and ${1}/{6}$ for testing. We label these six subsets of a given dataset according to their scanner domain as ($Train_1$, $Train_2$), ($Val_1$, $Val_2$), and ($Test_1$, $Test_2$), respectively.

\begin{figure*}[tp]
\includegraphics[width=\textwidth]{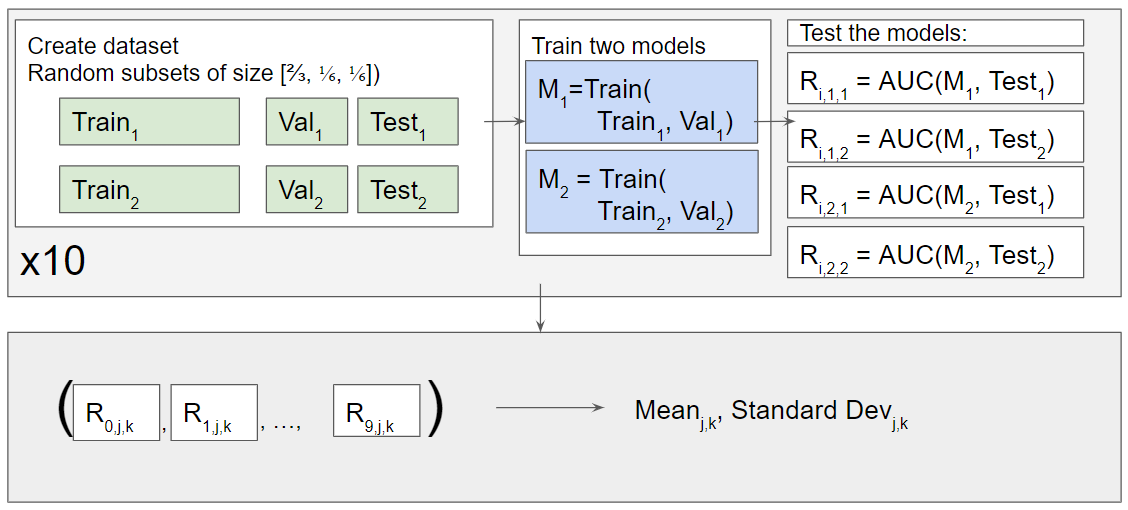}
\caption{\textbf{Seeded Training and Testing Procedure:} See Section \ref{sec:expdesign} for more details.}
\label{fig:expdesign}
\end{figure*}

Our study has two general types of tasks: the classification of 2D images (or slice images if taken from a 3D volume modality), and the classification of 3D crops from full 3D images. Both tasks have performance measured by the AUC (area under the receiver operating characteristic curve) \cite{bradley1997use} of the model's predictions on a given test set. Two separate models for each scanner type of a given dataset, denoted by $M_1$ and $M_2$, were trained and validated on their respective subsets ($Train_1$ with $Val_1$, $Train_2$ with $Val_2$). Each model was then tested on both the same manufacturer's test data and the other manufacturer's test data. To evaluate the variability and robustness of our results, we repeated all experiments ten times with different random seeds, reporting the results for a given model as the mean classification AUC with a standard deviation over all runs. Given this, we denote the performance of some model trained on data from scanner domain $j$ and tested on data from domain $k$ on the $i^{th}$ randomly-seeded run as $R_{i,j,k} := \mathrm{AUC}(M_j, Test_k)$, resulting in four average performance results per dataset (for each $j,k$ combination). This entire procedure is summarized in Figure \ref{fig:expdesign}.


\paragraph{Model and Training Details.} 2D image classification tasks are completed with the MRNet2D model adopted from \cite{Prakash_2021}, and 3D image tasks use the NoduleX model adopted from \cite{causey2018highly}. For the X-ray dataset we use DenseNet-121 \cite{huang2017densely}. Each model was trained for 60 epochs with a batch size of 50, and a learning rate of $0.0001$. All code will be publicly released upon paper acceptance to ensure the reproducibility of our results.


\subsection{Datasets, Preprocessing Methods, and Tasks}
The datasets we study include planar X-rays, magnetic resonance images (MRI), and computed tomography (CT) scans from various anatomical regions, each including images from two distinct scanner manufacturers, allowing for a comprehensive analysis of scanner domain shift phenomena. All datasets and their respective classification task labels and scanner domain pairs are provided in Table \ref{tab:datasets}. We applied $z$-score normalization to all images to ensure standardized pixel intensity distributions across all datasets, and in the following sections, we introduce each dataset and detail any additional pre-processing. We note that all datasets are publicly available to facilitate reproducibility.

\begin{table}[htbp]
\centering
\small
\begin{tabular}{@{}llllllll@{}}
\toprule
\textbf{Abbrev. Dataset Name} & \textbf{Full Dataset Name} & \textbf{Modality} & \textbf{Label} & \textbf{Manuf. 1} & \textbf{Manuf. 2}  \\
\midrule
MRI-Breast & Duke Breast Cancer MRI \cite{saha2018machine} & MRI & Tumor Stage $>2$ & GE  & Siemens  \\
MRI-Prostate & PI-CAI Challenge \cite{saha2023artificial} & MRI & Tumor & Siemens  & Other \\
CT-LIDC & Lung Image Database \cite{armato2011lung} & CT & High-Malignancy & GE  & Other  \\
CT-HNSCC & HNSCC CT \cite{grossberg2018imaging,elhalawani2017matched} & CT & Tumor & GE  & Other  \\
CT-Kidney & C4KC-KiTS \cite{heller2021state} & CT & Tumor & Siemens  & Other  \\
CT-Oropharyngeal & Radiomic Biomarkers \cite{kwan2018radiomic} & CT & HPV P16 Status & GE  & Toshiba  \\
Xray-Chest & Brazilian X-ray \cite{reis2022brax} & X-ray & Cardiomegaly & 3  & 5  \\
\bottomrule
\end{tabular}
\caption{All datasets used in this paper. ``\textbf{Label}'' is the label used for each dataset's classification task, and ``\textbf{Manuf.}'' stands for scanner manufacturer.}
\label{tab:datasets}
\end{table}

\paragraph{CT-Kidney \cite{heller2021state}:}

The dataset of kidney CTs was acquired at the University of Minnesota Medical Center between 2010 and 2018, from which we used the arterial view images. The diagnostic task is the binary classification of 2D slice images, distinguishing between those with or without a tumor, and the two domains are (1) images taken by Siemens-manufactured scanners and (2) images taken by other scanners.

\paragraph{CT-LIDC \cite{armato2011lung}:}

The LIDC / IDRI database contains thoracic CT scans collected from the Picture Archiving and Communications Systems (PACS) of seven participating academic institutions. For this dataset, our chosen task is to classify malignancies in 3D croppings of the full images. The images were cropped with a standardized 47 $\times$  47 $\times$  5 ROI enclosing any labeled lung nodules. The images were then given binary labels according to their level of nodule malignancy, either low or high. To ensure the robustness of our dataset, we only include nodules with extreme ratings of 1 (low) or 5 (high). The two domains correspond to GE scanners and other scanners, respectively.

\paragraph{CT-Oropharyngeal \cite{kwan2018radiomic}:}
This dataset contains CT scans of all non-metastatic p16-positive OPC patients treated with radiotherapy or chemoradiotherapy at a single institution between 2005 and 2010. Our binary classification task focuses on the presence of Human Papillomavirus (HPV). We extracted a 72x72x72 voxel ROI about the calculated centroid of any labeled tumor centered around this centroid, targeting the tumor's core area. These ROIs are then used for the classification task. Here, the two domains correspond to GE scanners and Toshiba scanners, respectively.

\paragraph{CT-HNSCC \cite{grossberg2018imaging}\cite{elhalawani2017matched}:}
This dataset consists of head and neck squamous cell carcinoma (HNSCC) CT scans from approximately 400 patients at the MD Anderson Cancer Center, with around 200-600 slice images per patient. The diagnostic task we evaluate is the binary classification of the presence of tumors in the 2D slices, based on whether a tumor mask is located in a particular slice or not. The two domains correspond to GE scanners and other scanners, respectively.

\paragraph{MRI-Prostate \cite{saha2023artificial}:}

This dataset contains prostate MRI scans of 1,476 patients, acquired between 2012-2021, in three centers (Radboud University Medical Center, University Medical Center Groningen, Ziekenhuis Groep Twente) based in The Netherlands. We perform a binary slice classification task for the presence of any tumor in a given slice. For this dataset, the two domains correspond to Siemens scanners and other scanners, respectively.

\paragraph{MRI-Breast \cite{saha2018machine}:}

This dataset is a retrospective collection of breast cancer MRIs from a single institution (Duke University) over a decade. We tested binary classification based on the stage of the breast cancer, specifically whether the cancer stage is greater than 2 or not. We extracted a 3D ROI with dimensions 47 $\times$ 47 $\times$ 5 centered around any labeled tumor (according to the center of the provided bounding boxes), which is then used to perform the classification. In this case, the two domains correspond to GE scanners and Siemens scanners.

\paragraph{Xray-Chest \cite{reis2022brax}:}
This dataset contains 24,959 chest radiographs of patients who presented at a large general hospital in Brazil. The task involves classifying images based on the absence or presence of various cardiomegaly-related conditions. In this dataset, the domains correspond to two anonymized scanner manufacturers, ``3'' and ``5'' in their paper.

\section{Results}




\subsection{Effects of Scanner Domain Shift}
\label{sec:shiftexp}
\begin{figure*}[tp]
\includegraphics[width=\textwidth]{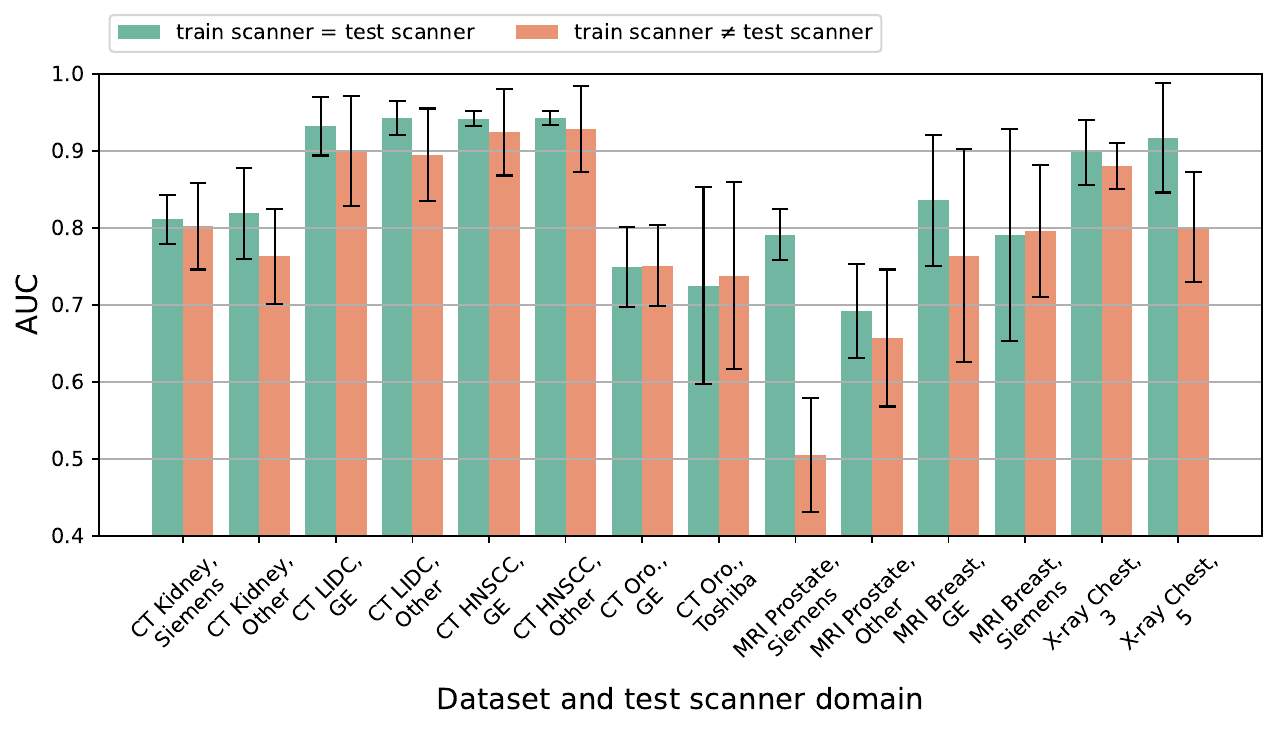}
\caption{Visualization of the change in network performance due to the test scanner domain shifting away from the training scanner domain (see Table \ref{tab:shift} for reference). Each pair of bars is for a single modality and training set scanner manufacturer.}
\label{Barplot}
\end{figure*}

In Table \ref{tab:shift} and Figure \ref{Barplot} we show the AUC performance $R_{i,j,k} := \mathrm{AUC}(M_j, Test_k)$ of the model trained in scanner domain $j$ and tested in domain $k$, averaged over all ten randomly-seeded runs indexed by $i$, for each dataset.

\begin{table}[htbp]
\centering
\small
\begin{tabular}{@{}lllllcll@{}}
\toprule
\textbf{Modality} & \textbf{Dataset} & \textbf{\tworow{Training scanner}{manufacturer}} & \textbf{\tworow{Test performance:}{same manuf.}} & \textbf{\tworow{Test performance:}{different manuf.}} & \textbf{$\Delta$}  \\
\midrule
CT                & Kidney           & Other                & 0.819 $\pm$ 0.059               & 0.763 $\pm$ 0.062                   & -0.056        \\
CT                & LIDC             & GE                   & 0.932 $\pm$ 0.038              & 0.900 $\pm$ 0.071                   & -0.032         \\
CT                & LIDC             & Other                & 0.943 $\pm$ 0.022               & 0.895 $\pm$ 0.060                   & -0.048       \\
CT                & HNSCC            & GE                  & 0.942 $\pm$ 0.010               & 0.924 $\pm$ 0.056                   & -0.018       \\
CT                & HNSCC            & Other               & 0.943 $\pm$ 0.009               & 0.928 $\pm$ 0.056                   & -0.015        \\
CT                & Oropharyngeal    & GE                   & 0.749 $\pm$ 0.052               & 0.751 $\pm$ 0.053                   & +0.002        \\
CT                & Oropharyngeal    & Toshiba              & 0.725 $\pm$ 0.128               & 0.738 $\pm$ 0.121                  & +0.013        \\
MRI               & Prostate         & Siemens              & 0.791 $\pm$ 0.033               & 0.505 $\pm$ 0.074                   & -0.286         \\
MRI               & Prostate         & Other                & 0.692 $\pm$ 0.061               & 0.657 $\pm$ 0.089                   & -0.035         \\
MRI               & Breast           & GE              & 0.836 $\pm$ 0.085               & 0.764 $\pm$ 0.138                   & -0.072         \\
MRI               & Breast           & Siemens                   & 0.791 $\pm$ 0.138               & 0.796 $\pm$ 0.086                   & +0.005         \\

X-ray             & Chest              & 3                  & 0.898  $\pm$ 0.042                              & 0.880 $\pm$ 0.030 & -0.018 \\
X-ray             & Chest              & 5                  & 0.917  $\pm$ 0.071                              & 0.801 $\pm$ 0.071 &  -0.116 \\

\bottomrule
\end{tabular}
\caption{\textbf{Effect of scanner domain shift on model performance (AUC).} ``Same/different manuf.'' refers to the test set of images from either the same scanner manufacturer that the neural network was training on, or from a different scanner. Confidence intervals are standard deviations over all $10$ runs (Fig. \ref{fig:expdesign}). $\Delta$ is the difference between the two performances, showing the effect of scanner domain shift.}
\label{tab:shift}
\end{table}

First, we see that unsurprisingly, models generally achieved higher accuracy when their test set was of the same domain/distribution as their training set. An interesting exception was observed in the results for the CT-HNSCC and CT-Oropharyngeal datasets, where the AUC scores were similar for all experiments for both matched and mismatched training and testing domains. This implies that the visual features in the images that are needed for the task are similar for different scanner types, and so those from one domain transfer to the other. 

\section*{Discussion}

This study comprehensively examined the impact of scanner domain shift on the performance of deep neural networks in medical image analysis. Overall, our findings align with previous research indicating the sensitivity of deep learning models to variations in image features, even subtle ones. We hypothesize that these variations in visual features are likely due to differences in scanner parameters and post-processing techniques between imaging sites.
\subsubsection*{The Dependence of Scanner Domain Shift on Modality}

\begin{figure*}[tp]
\centering
\includegraphics[width=0.3\textwidth]{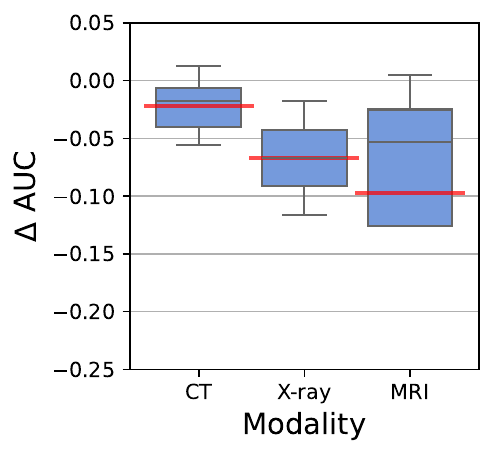}
\caption{Box-and-whisker plot of the change in neural network classification performance due to scanner domain shift across different modalities (see Table \ref{tab:shift}). Average values shown with red lines.}
\label{fig:scannershift_barplot}
\end{figure*}

Notably, the scanner domain shift results (Table \ref{tab:shift}) tell of an interesting pattern: the typical severity of domain shift changes depending on the imaging modality. Here, we will use $\Delta$ to denote the change from in-domain to out-of-domain test set AUC.
We see that domain shift is typically most severe for MRI tasks ($\Delta = -0.097$ on average), moderate for X-ray ($\Delta = -0.067$), and quite small for CT ($\Delta = - 0.02$), as summarized in Fig. \ref{fig:scannershift_barplot}. We hypothesize that this varying severity of domain shift is dependent on both how much scanner image acquisition parameters and proprietary post-processing methods usually differ between imaging settings/domains for a given scanner type, and the strength of the effect of these factors on image appearance (excluding potential confounding effects such as biases in patient population). This is because domains with different scanner parameters and/or post-processing will have the effect of differing distributions of generated image features which the network learns for the downstream task, so that certain features learned from one domain may not be present in the other, resulting in a loss of performance.

While it is infeasible in generality to directly quantify how differences in imaging parameters/effects between domains for a given modality result in differences in acquired image features (and then therefore downstream network task performance), we can at least explore the nature of these parameters to hypothesize why some modalities may be more susceptible to domain shift than others, in terms of their acquisition system which may vary between scanner domains. Note that proprietary post-processing unique to a given scanner model/manufacturer can also affect image appearance, but by being proprietary, these effects are unknown.

MRI has many different scanner parameters that will directly affect image appearance, such as echo time (TE), repetition time (TR) and flip angle \cite{nitz1999contrast}, which all differ in distribution between GE and Siemens scans for the breast MRI dataset \cite{konz2023reverse}. Changing any of these parameters will affect the visible contrast of different types of tissue in accordance with their intrinsic $T_1$ and/or $T_2$ values. Indeed, for breast MRI, previous research \cite{saha2017effects} has suggested that variations in scanner parameters affect the image/radiomic features of fibroglandular tissue in breast MRI, potentially impacting the robustness of models trained on these features. This insight may explain the scanner-specific performance disparities observed in the MRI dataset, particularly in features related to tumor detection.

X-ray is also affected by imaging parameters and other effects which can result in image features that differ between domains. Examples of these parameters and effects include the spectrum, energy/dosage, and beam angle of the radiation, as well as potential measurement noise, according to an extension of the Beer-Lambert law \cite{kilim2022physical}. Finally, CT images may be similarly affected by imaging parameters such as slice thickness, pixel size, and dosage \cite{huang2021impact}, although these are much more standardized, resulting in less severe domain shift effects.


\subsection*{Other Results and Future Work}

Additionally, we found that adding simple Gaussian noise to images in training did not assist with performance in other domains, indicating that more complex solutions to prevent in-domain overfitting and encourage out-of-domain generalization \cite{wang2022generalizing} are necessary.

Future studies could expand on ours in various ways. While we chose to have breadth of datasets and modalities, further studies could explore the presence of the phenomena for additional network architectures and training set sizes. Additionally, the effect of scanner shift on different diagnostic tasks could be evaluated, such as other classification tasks, or other tasks altogether such as regression, object detection, or semantic segmentation. Pixel-level localization tasks such as detection and segmentation may be more susceptible to domain shift due to them (1) often being more challenging than image-level classification and (2) being dense prediction tasks that may have more capacity for overfitting than image-level prediction tasks such as classification.

\section*{Conclusion}

Our findings emphasize the importance of accounting for scanner variability when developing and validating deep learning models for medical image diagnosis, and highlight a potential challenge in deploying these models across different clinical settings where scanner heterogeneity is common. Further research into methods for retroactively mitigating the impact of scanner-specific factors on image appearance may be critical to enhance the generalizability and clinical applicability of deep learning methods for radiology.

\section*{Acknowledgments}
 Research reported in this publication was supported by the National Institute Of Biomedical Imaging And Bioengineering of the National Institutes of Health under Award Number R01EB031575. The content is solely the responsibility of the authors and does not necessarily represent the official views of the National Institutes of Health.

\bibliographystyle{plain}  
\bibliography{references}  

\begin{thebibliography}{10}

\bibitem{albadawy2018deep}
Ehab~A AlBadawy, Ashirbani Saha, and Maciej~A Mazurowski.
\newblock Deep learning for segmentation of brain tumors: Impact of
  cross-institutional training and testing.
\newblock {\em Medical physics}, 45(3):1150--1158, 2018.

\bibitem{armato2011lung}
Samuel~G Armato~III, Geoffrey McLennan, Luc Bidaut, Michael~F McNitt-Gray,
  Charles~R Meyer, Anthony~P Reeves, Binsheng Zhao, Denise~R Aberle, Claudia~I
  Henschke, Eric~A Hoffman, et~al.
\newblock The lung image database consortium (lidc) and image database resource
  initiative (idri): a completed reference database of lung nodules on ct
  scans.
\newblock {\em Medical physics}, 38(2):915--931, 2011.

\bibitem{bradley1997use}
Andrew~P Bradley.
\newblock The use of the area under the roc curve in the evaluation of machine
  learning algorithms.
\newblock {\em Pattern recognition}, 30(7):1145--1159, 1997.

\bibitem{cao2023deep}
Shixing Cao, Nicholas Konz, James Duncan, and Maciej~A Mazurowski.
\newblock Deep learning for breast mri style transfer with limited training
  data.
\newblock {\em Journal of Digital Imaging}, 36(2):666--678, 2023.

\bibitem{causey2018highly}
Jason Causey, Junyu Zhang, Shiqian Ma, Bo~Jiang, Jake Qualls, David~G. Politte,
  Fred Prior, Shuzhong Zhang, and Xiuzhen Huang.
\newblock Highly accurate model for prediction of lung nodule malignancy with
  ct scans, 2018.

\bibitem{celard2023survey}
Pedro Celard, Eva~Lorenzo Iglesias, Jos{\'e}~Manuel Sorribes-Fdez, Rub{\'e}n
  Romero, A~Seara Vieira, and Lourdes Borrajo.
\newblock A survey on deep learning applied to medical images: from simple
  artificial neural networks to generative models.
\newblock {\em Neural Computing and Applications}, 35(3):2291--2323, 2023.

\bibitem{chan2020deep}
Heang-Ping Chan, Ravi~K Samala, Lubomir~M Hadjiiski, and Chuan Zhou.
\newblock Deep learning in medical image analysis.
\newblock {\em Deep learning in medical image analysis: challenges and
  applications}, pages 3--21, 2020.

\bibitem{elhalawani2017matched}
Hesham Elhalawani, Abdallah~SR Mohamed, Aubrey~L White, James Zafereo, Andrew~J
  Wong, Joel~E Berends, Shady AboHashem, Bowman Williams, Jeremy~M Aymard,
  Aasheesh Kanwar, et~al.
\newblock Matched computed tomography segmentation and demographic data for
  oropharyngeal cancer radiomics challenges.
\newblock {\em Scientific data}, 4:170077, 2017.

\bibitem{geirhos2018imagenet}
Robert Geirhos, Patricia Rubisch, Claudio Michaelis, Matthias Bethge, Felix~A
  Wichmann, and Wieland Brendel.
\newblock Imagenet-trained cnns are biased towards texture; increasing shape
  bias improves accuracy and robustness.
\newblock In {\em International Conference on Learning Representations}, 2018.

\bibitem{grossberg2018imaging}
Aaron~J Grossberg, Abdallah~SR Mohamed, Hesham Elhalawani, William~C Bennett,
  Kirk~E Smith, Tracy~S Nolan, Bowman Williams, Sasikarn Chamchod, Jolien
  Heukelom, Michael~E Kantor, et~al.
\newblock Imaging and clinical data archive for head and neck squamous cell
  carcinoma patients treated with radiotherapy.
\newblock {\em Scientific data}, 5(1):1--10, 2018.

\bibitem{guan2021domain}
Hao Guan and Mingxia Liu.
\newblock Domain adaptation for medical image analysis: a survey.
\newblock {\em IEEE Transactions on Biomedical Engineering}, 69(3):1173--1185,
  2021.

\bibitem{heller2021state}
Nicholas Heller, Fabian Isensee, Klaus~H Maier-Hein, Xiaoshuai Hou, Chunmei
  Xie, Fengyi Li, Yang Nan, Guangrui Mu, Zhiyong Lin, Miofei Han, et~al.
\newblock The state of the art in kidney and kidney tumor segmentation in
  contrast-enhanced ct imaging: Results of the kits19 challenge.
\newblock {\em Medical image analysis}, 67:101821, 2021.

\bibitem{huang2017densely}
Gao Huang, Zhuang Liu, Laurens Van Der~Maaten, and Kilian~Q Weinberger.
\newblock Densely connected convolutional networks.
\newblock In {\em Proceedings of the IEEE conference on computer vision and
  pattern recognition}, pages 4700--4708, 2017.

\bibitem{huang2021impact}
Kai Huang, Dong~Joo Rhee, Rachel Ger, Rick Layman, Jinzhong Yang, Carlos~E
  Cardenas, and Laurence~E Court.
\newblock Impact of slice thickness, pixel size, and ct dose on the performance
  of automatic contouring algorithms.
\newblock {\em Journal of applied clinical medical physics}, 22(5):168--174,
  2021.

\bibitem{kilim2022physical}
Oz~Kilim, Alex Olar, Tam{\'a}s Jo{\'o}, Tam{\'a}s Palicz, P{\'e}ter Pollner,
  and Istv{\'a}n Csabai.
\newblock Physical imaging parameter variation drives domain shift.
\newblock {\em Scientific Reports}, 12(1):21302, 2022.

\bibitem{konz2023reverse}
Nicholas Konz and Maciej~A Mazurowski.
\newblock Reverse engineering breast mris: Predicting acquisition parameters
  directly from images.
\newblock In {\em Medical Imaging with Deep Learning}, 2023.

\bibitem{kwan2018radiomic}
Jennifer Yin~Yee Kwan, Jie Su, Shao~Hui Huang, Laleh~S Ghoraie, Wei Xu, Biu
  Chan, Kenneth~W Yip, Meredith Giuliani, Andrew Bayley, John Kim, et~al.
\newblock Radiomic biomarkers to refine risk models for distant metastasis in
  hpv-related oropharyngeal carcinoma.
\newblock {\em International Journal of Radiation Oncology* Biology* Physics},
  102(4):1107--1116, 2018.

\bibitem{litjens2017survey}
Geert Litjens, Thijs Kooi, Babak~Ehteshami Bejnordi, Arnaud Arindra~Adiyoso
  Setio, Francesco Ciompi, Mohsen Ghafoorian, Jeroen~Awm Van Der~Laak, Bram
  Van~Ginneken, and Clara~I S{\'a}nchez.
\newblock A survey on deep learning in medical image analysis.
\newblock {\em Medical image analysis}, 42:60--88, 2017.

\bibitem{madry2018towards}
Aleksander Madry, Aleksandar Makelov, Ludwig Schmidt, Dimitris Tsipras, and
  Adrian Vladu.
\newblock Towards deep learning models resistant to adversarial attacks.
\newblock In {\em International Conference on Learning Representations}, 2018.

\bibitem{maartensson2020reliability}
Gustav M{\aa}rtensson, Daniel Ferreira, Tobias Granberg, Lena Cavallin, Ketil
  Oppedal, Alessandro Padovani, Irena Rektorova, Laura Bonanni, Matteo Pardini,
  Milica~G Kramberger, et~al.
\newblock The reliability of a deep learning model in clinical
  out-of-distribution mri data: a multicohort study.
\newblock {\em Medical Image Analysis}, 66:101714, 2020.

\bibitem{nitz1999contrast}
Wolfgang~R Nitz and P~Reimer.
\newblock Contrast mechanisms in mr imaging.
\newblock {\em European radiology}, 9:1032--1046, 1999.

\bibitem{Prakash_2021}
Jatin Prakash.
\newblock Stanford mrnet challenge: Classifying knee mris: Learnopencv, May
  2021.

\bibitem{reis2022brax}
Eduardo~P Reis, Joselisa~PQ de~Paiva, Maria~CB da~Silva, Guilherme~AS Ribeiro,
  Victor~F Paiva, Lucas Bulgarelli, Henrique~MH Lee, Paulo~V Santos, Vanessa~M
  Brito, Lucas~TW Amaral, et~al.
\newblock Brax, brazilian labeled chest x-ray dataset.
\newblock {\em Scientific Data}, 9(1):487, 2022.

\bibitem{saha2023artificial}
Anindo Saha, Joeran Bosma, Jasper Twilt, Bram van Ginneken, Derya Yakar,
  Mattijs Elschot, Jeroen Veltman, Jurgen F{\"u}tterer, Maarten de~Rooij,
  et~al.
\newblock Artificial intelligence and radiologists at prostate cancer detection
  in mri—the pi-cai challenge.
\newblock In {\em Medical Imaging with Deep Learning, short paper track}, 2023.

\bibitem{saha2018machine}
Ashirbani Saha, Michael~R Harowicz, Lars~J Grimm, Connie~E Kim, Sujata~V Ghate,
  Ruth Walsh, and Maciej~A Mazurowski.
\newblock A machine learning approach to radiogenomics of breast cancer: a
  study of 922 subjects and 529 dce-mri features.
\newblock {\em British journal of cancer}, 119(4):508--516, 2018.

\bibitem{saha2017effects}
Ashirbani Saha, Xiaozhi Yu, Dushyant Sahoo, and Maciej~A Mazurowski.
\newblock Effects of mri scanner parameters on breast cancer radiomics.
\newblock {\em Expert systems with applications}, 87:384--391, 2017.

\bibitem{szegedy2013intriguing}
Christian Szegedy, Wojciech Zaremba, Ilya Sutskever, Joan Bruna, Dumitru Erhan,
  Ian Goodfellow, and Rob Fergus.
\newblock Intriguing properties of neural networks.
\newblock {\em arXiv preprint arXiv:1312.6199}, 2013.

\bibitem{wang2022generalizing}
Jindong Wang, Cuiling Lan, Chang Liu, Yidong Ouyang, Tao Qin, Wang Lu, Yiqiang
  Chen, Wenjun Zeng, and Philip Yu.
\newblock Generalizing to unseen domains: A survey on domain generalization.
\newblock {\em IEEE Transactions on Knowledge and Data Engineering}, 2022.

\bibitem{wang2022embracing}
Rongguang Wang, Pratik Chaudhari, and Christos Davatzikos.
\newblock Embracing the disharmony in medical imaging: A simple and effective
  framework for domain adaptation.
\newblock {\em Medical image analysis}, 76:102309, 2022.

\bibitem{wang2020inconsistent}
Xiaoqin Wang, Gongbo Liang, Yu~Zhang, Hunter Blanton, Zachary Bessinger, and
  Nathan Jacobs.
\newblock Inconsistent performance of deep learning models on mammogram
  classification.
\newblock {\em Journal of the American College of Radiology}, 17(6):796--803,
  2020.

\bibitem{yao2020strong}
Li~Yao, Jordan Prosky, Ben Covington, and Kevin Lyman.
\newblock A strong baseline for domain adaptation and generalization in medical
  imaging.
\newblock In {\em Medical Imaging with Deep Learning - Extended Abstract
  Track}, 2019.

\bibitem{cyclegan}
Jun-Yan Zhu, Taesung Park, Phillip Isola, and Alexei~A Efros.
\newblock Unpaired image-to-image translation using cycle-consistent
  adversarial networks.
\newblock In {\em Proceedings of the IEEE international conference on computer
  vision}, pages 2223--2232, 2017.

\bibitem{pmid19928111}
R.~M. Zur, Y.~Jiang, L.~L. Pesce, and K.~Drukker.
\newblock {{N}oise injection for training artificial neural networks: a
  comparison with weight decay and early stopping}.
\newblock {\em Med Phys}, 36(10):4810--4818, Oct 2009.

\end{thebibliography}

\newpage
\section*{Supplementary Materials}
\appendix

\section{Additional Experiments}
\label{app:additionalexp}
\subsection{How does additional training data from the target domain help domain generalization?}
\label{app:mixing}
Given that almost all datasets suffer from some level of performance drop when the scanner domain of the test set differs from that of the training set (Table \ref{tab:shift}), a logical next exploration would be to explore this continuum of performance drop between these two domains. To do so, we next evaluated how test set performance is affected by the training set having varying amounts of target domain images inserted. Specifically, we move images randomly from the target domain's training set to the original source domain training set such that a fraction $f\in [0.1,0.2,0.3,0.4,0.5]$ of the training set is comprised of the target domain images. Once the model is trained on this mixed training set, we evaluate it on the usual target domain test set. We repeated this procedure three times with different random seeds, for each of the three datasets of MRI-Prostate, CT-LIDC, and CT-Kidney. When the target domain training dataset did not have enough samples to reach the desired $f$, we duplicated samples until we reached the specified mixin amount for the training set.

\begin{figure*}[htpb!]
\centering
\includegraphics[width=0.49\linewidth]{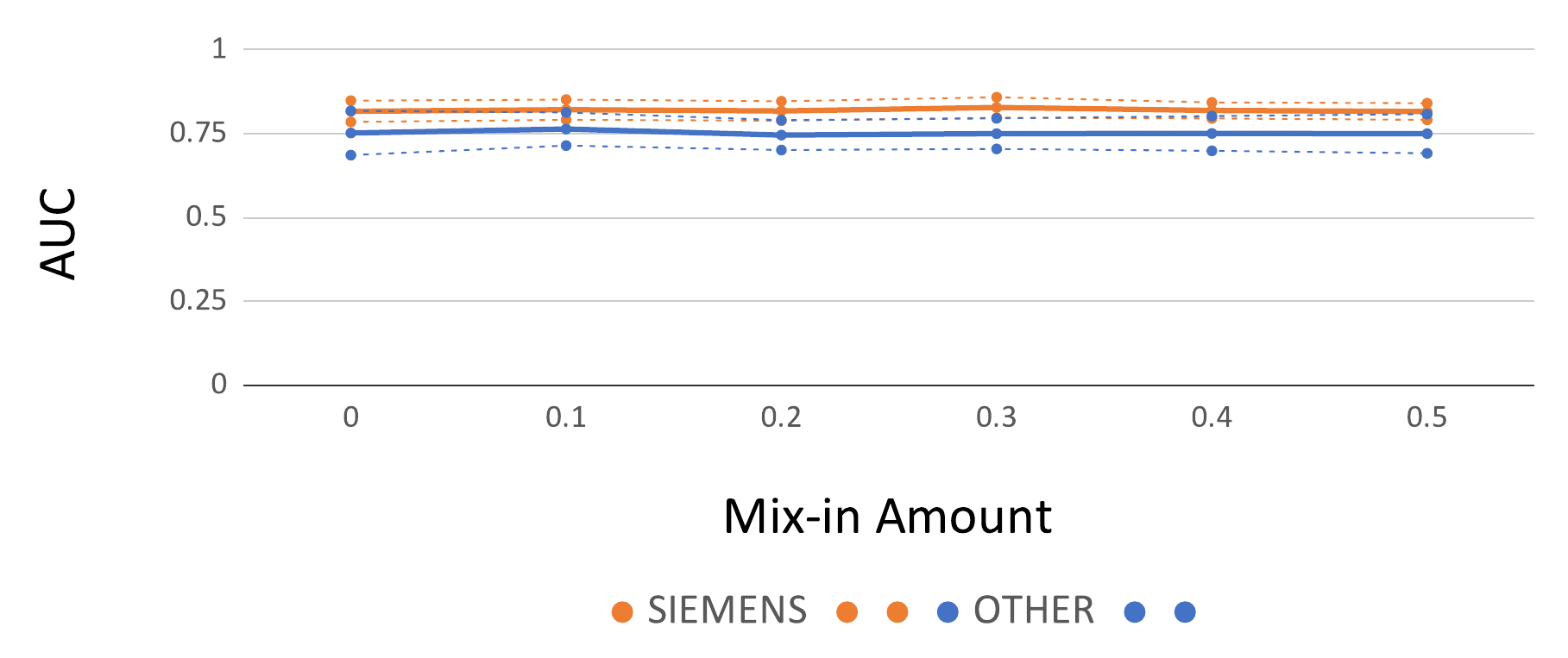}
\includegraphics[width=0.49\linewidth]{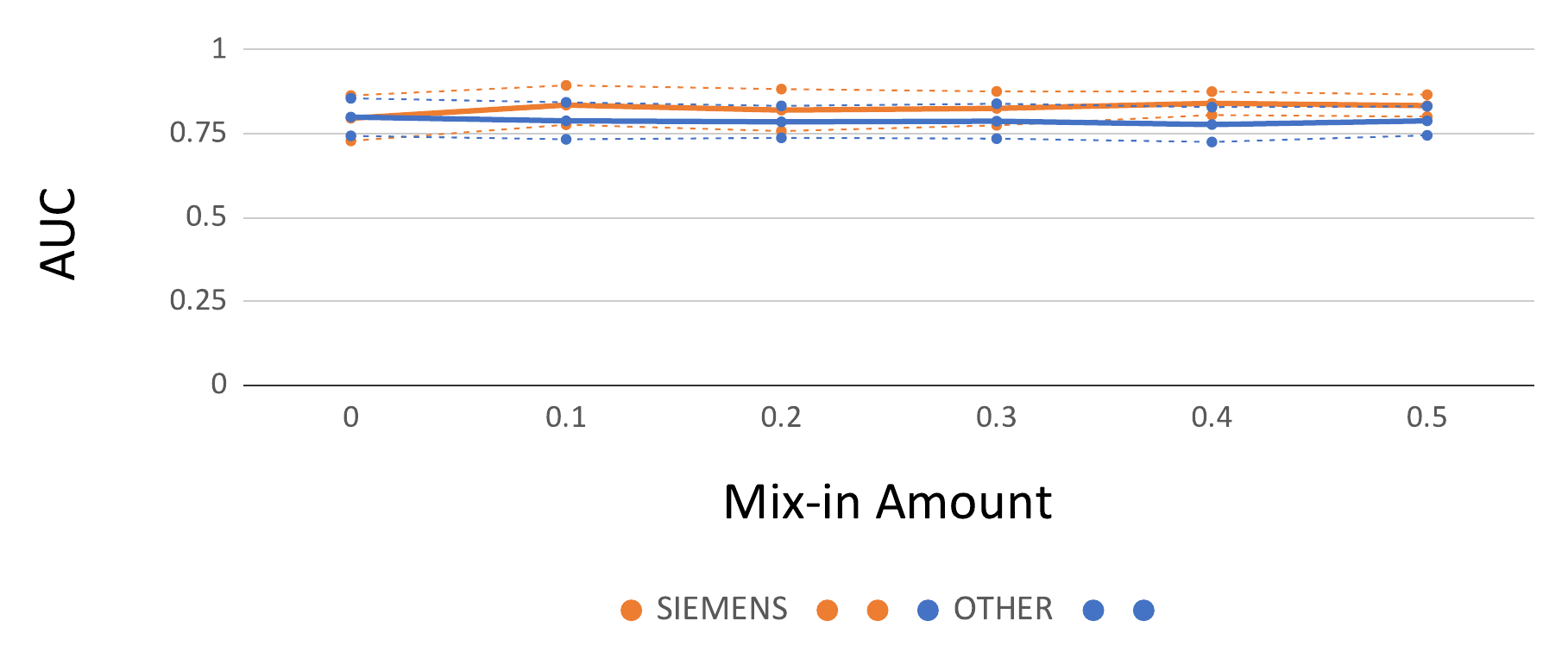}
\includegraphics[width=0.49\linewidth]{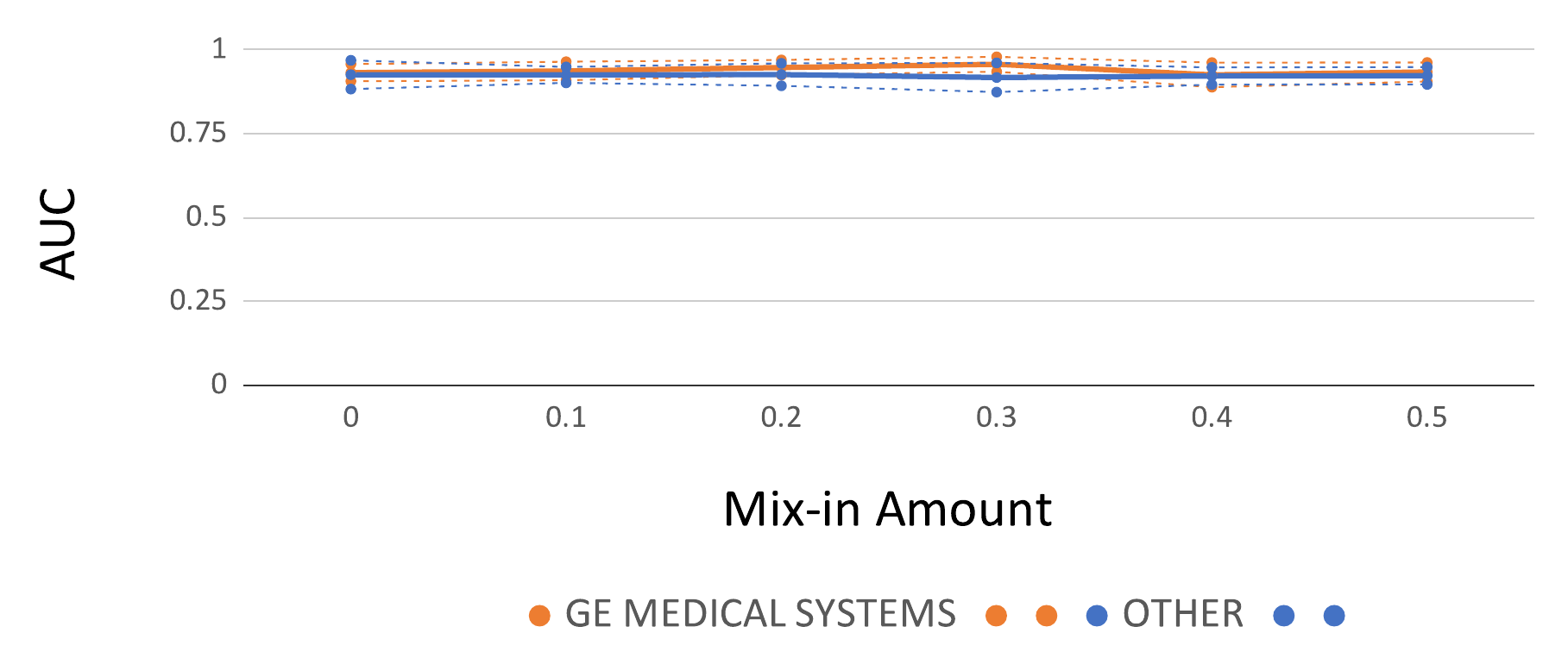}
\includegraphics[width=0.49\linewidth]{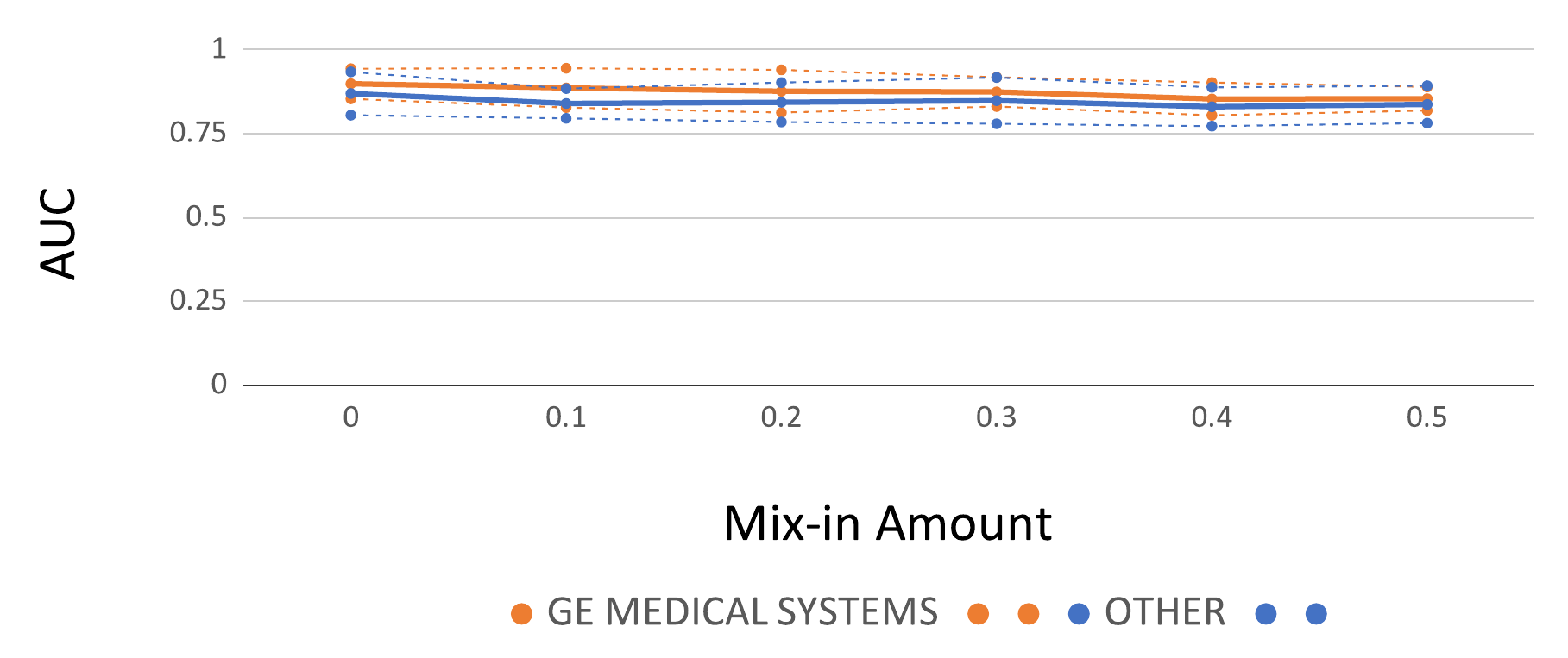}
\includegraphics[width=0.49\linewidth]{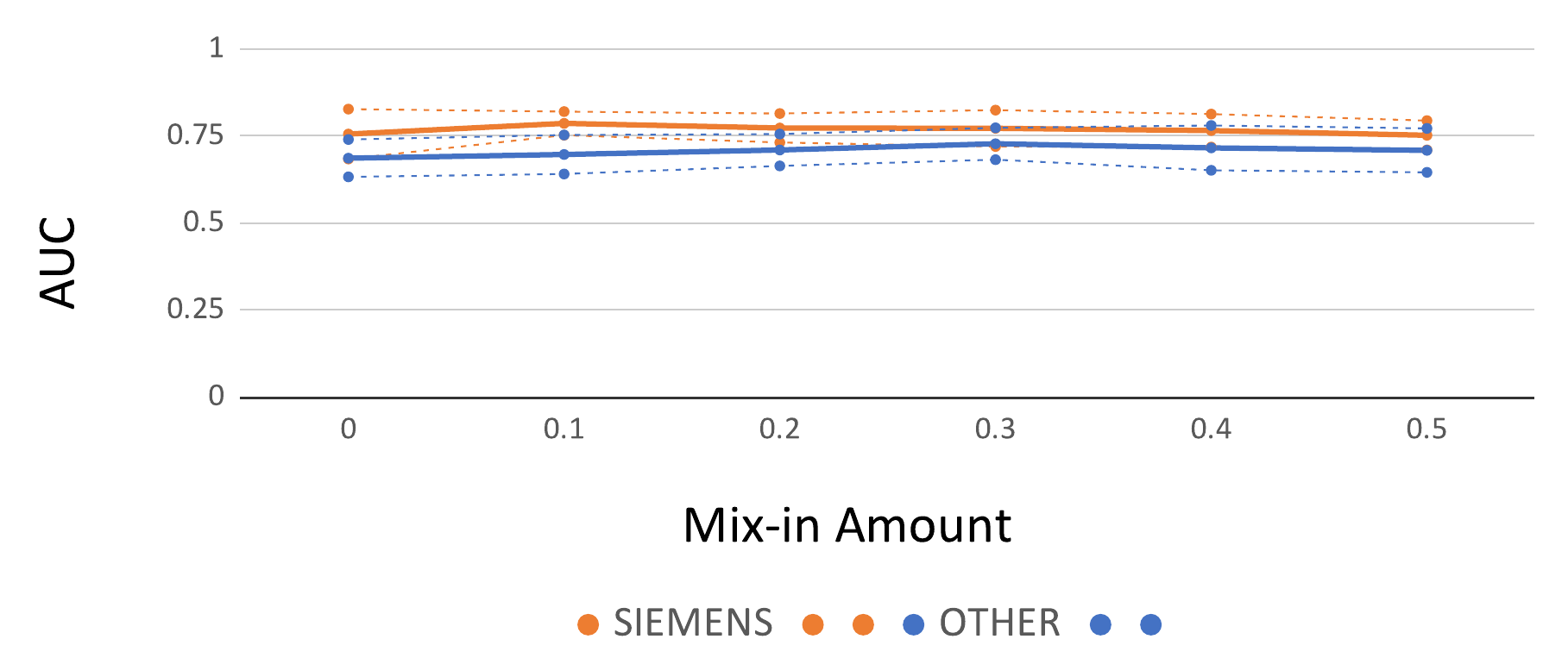}
\includegraphics[width=0.49\linewidth]{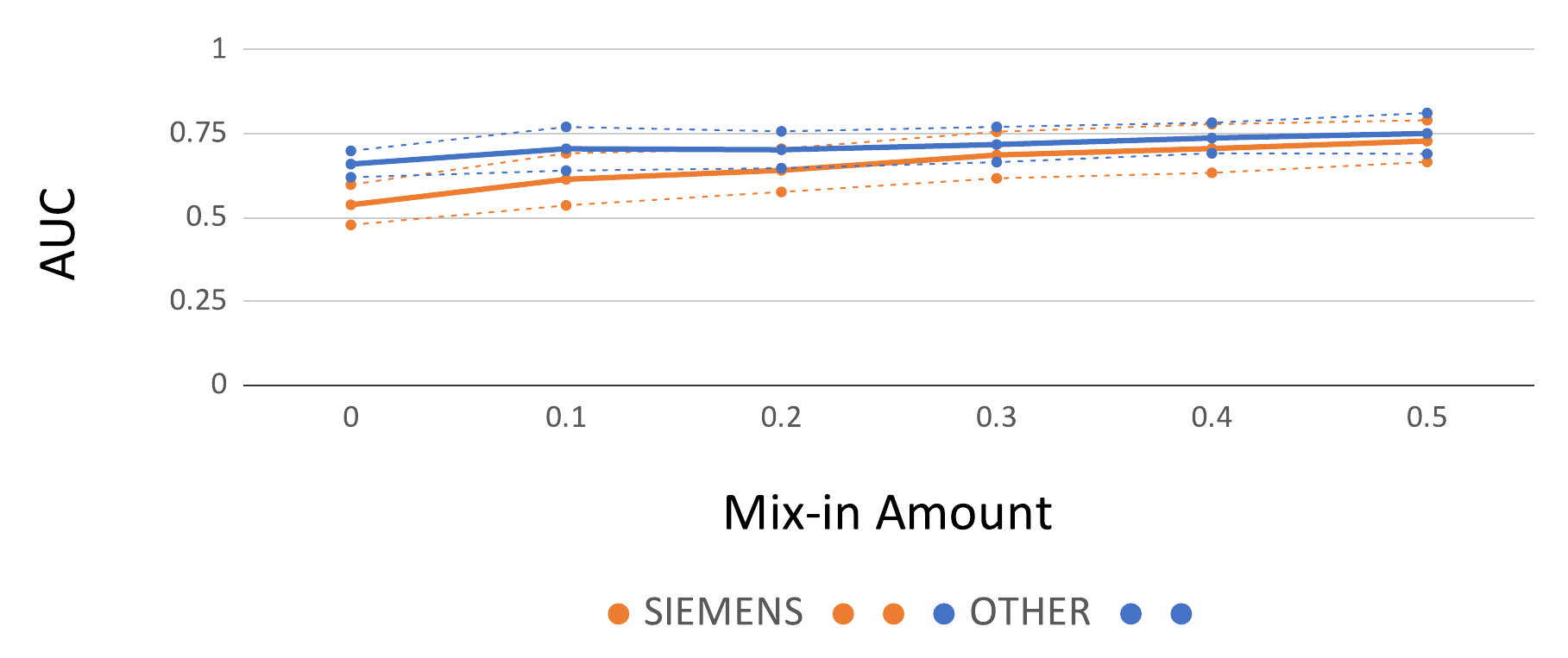}
\caption{Scanner domain mixing experiment results for \textbf{(a) CT-Kidney} with Siemens and Other training domains on the left and right, respectively; \textbf{(b) CT-LIDC} with GE and Other training domains on the left and right; and \textbf{(c) MRI-Prostate} with Siemens and Other training domains. Different colors correspond to different scanner domains for the test set. Confidence intervals (standard deviation over all runs) shown with dotted lines.}
\label{fig:mixin}
\end{figure*}

The results of these scanner mixing experiments (test set AUC as a function of mixing fraction $f$) are presented in Figure \ref{fig:mixin}. We see that scanner mixing for CT tasks results in little change in performance, while the effect is more noticeable for MRI. This is reasonable given that domain shift was typically small for CT, yet larger for MRI (Section \ref{sec:shiftexp}), likely due to differing similarity/overlap of features between source and target domains for the two modalities. For MRI in particular, we see an asymmetry in the degree of domain shift depending on the training domain: models trained on solely Siemens images generalize well to the target domain, such that mixing has little effect, but models trained on ``Other'' domain images don't generalize as well, such that mixing in images from the target domain provided noticeable benefit. This is in turn due to features learned from one domain generalizing well to the other, but not vice versa.

\subsection{Does adding noise to training images assist domain generalization?}
\label{app:noise}

Adding noise to training data has been shown to help deep learning models reduce overfitting and generalize better to unseen data \cite{pmid19928111} (for example, adversarial attack-resistant training \cite{madry2018towards} can be considered as a specialization of this idea). Moreover, unlike in general natural image computer vision where images from different domains often have clear, noticeable differences in visual features (\eg, photographs vs. paintings or cartoon sketches \cite{cyclegan}), images from different scanner domains in medical images may not even appear visibly distinct (such as for the breast MRI dataset \cite{cao2023deep}), implying that differing visual features may just be fine-grained characteristics such as noise and slight changes in texture. As such, simply adding basic noise to one of our model's training data could potentially result in better generalization to other domains, which we will explore in this section.

For these experiments, we evaluated adding random Gaussian noise of different levels to training set images. For each training set image $x\in\mathbb{R}^n$, we add $\alpha \epsilon$ to $x$ with $\epsilon \sim \mathcal{N}(0, I_n)$ and $\alpha\in[1,2,3,4,5,7,10,15,30,50]$. As in the scanner mixing experiments (Appendix \ref{app:mixing}), we evaluate on the MRI-Prostate, CT-LIDC, and CT-Kidney datasets, and repeat all experiments at each noise level three times.

\begin{figure*}[htpb!]
\centering
\includegraphics[width=0.49\linewidth]{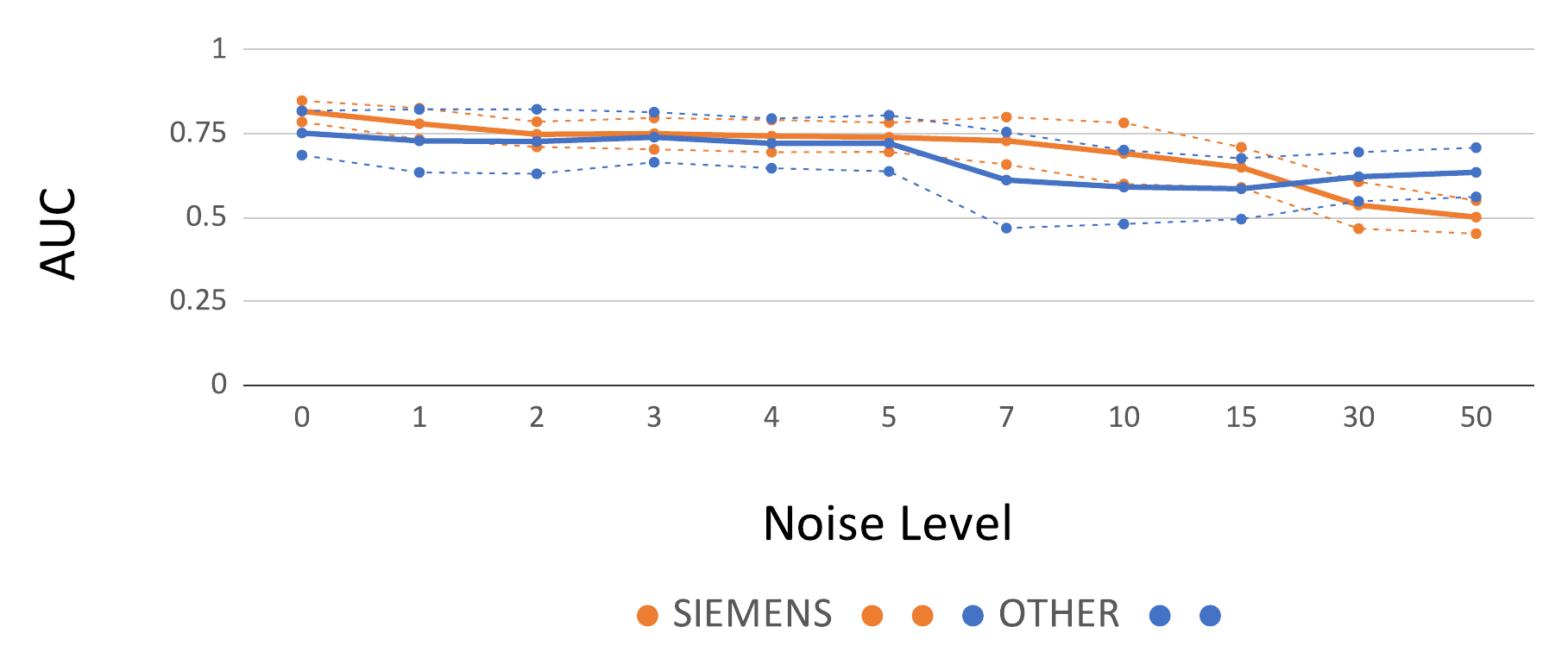}
\includegraphics[width=0.49\linewidth]{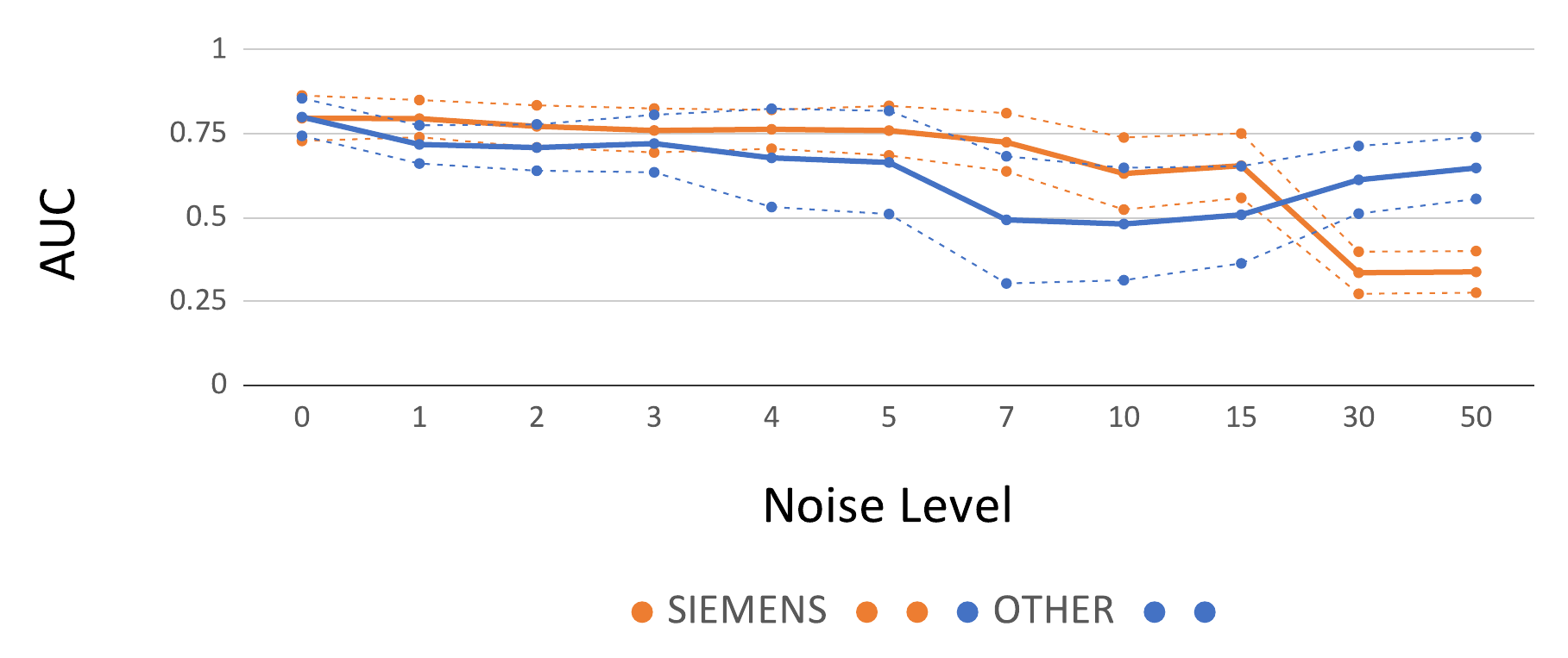}
\includegraphics[width=0.49\linewidth]{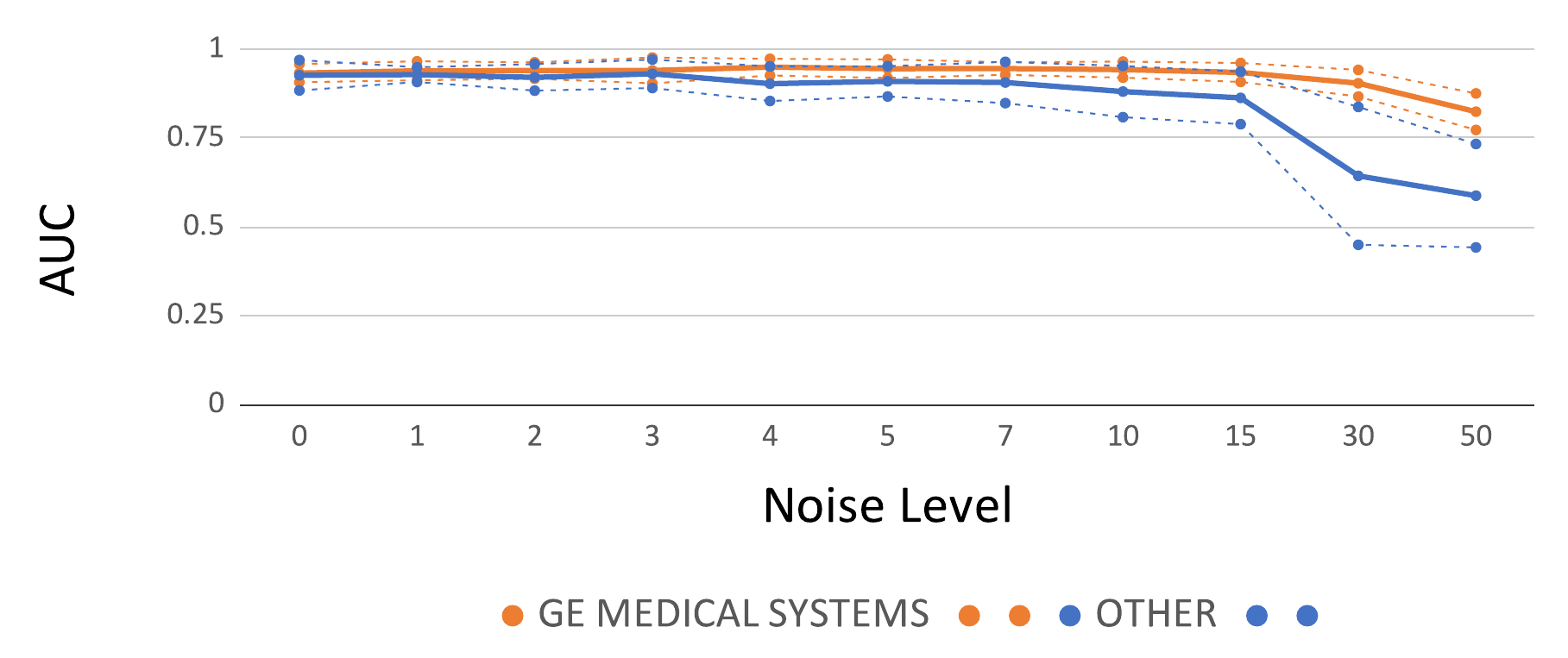}
\includegraphics[width=0.49\linewidth]{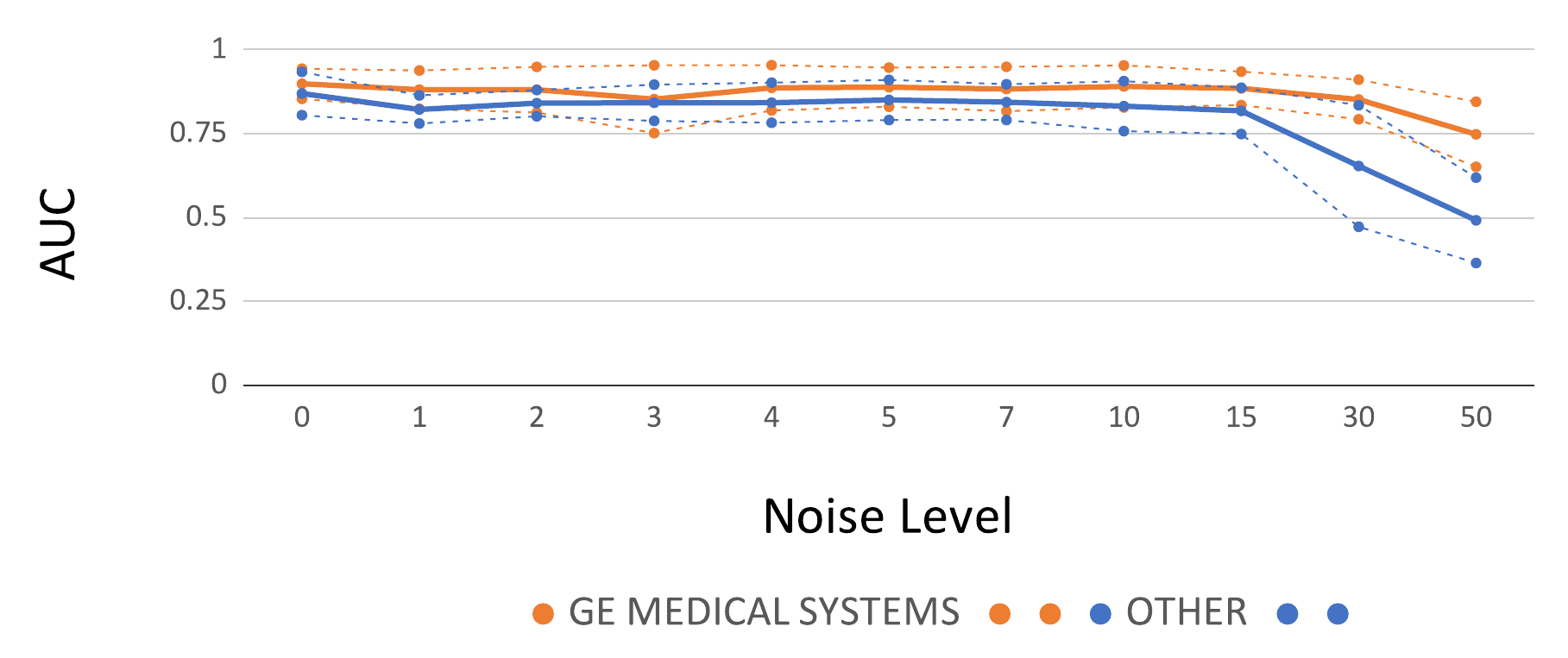}
\includegraphics[width=0.49\linewidth]{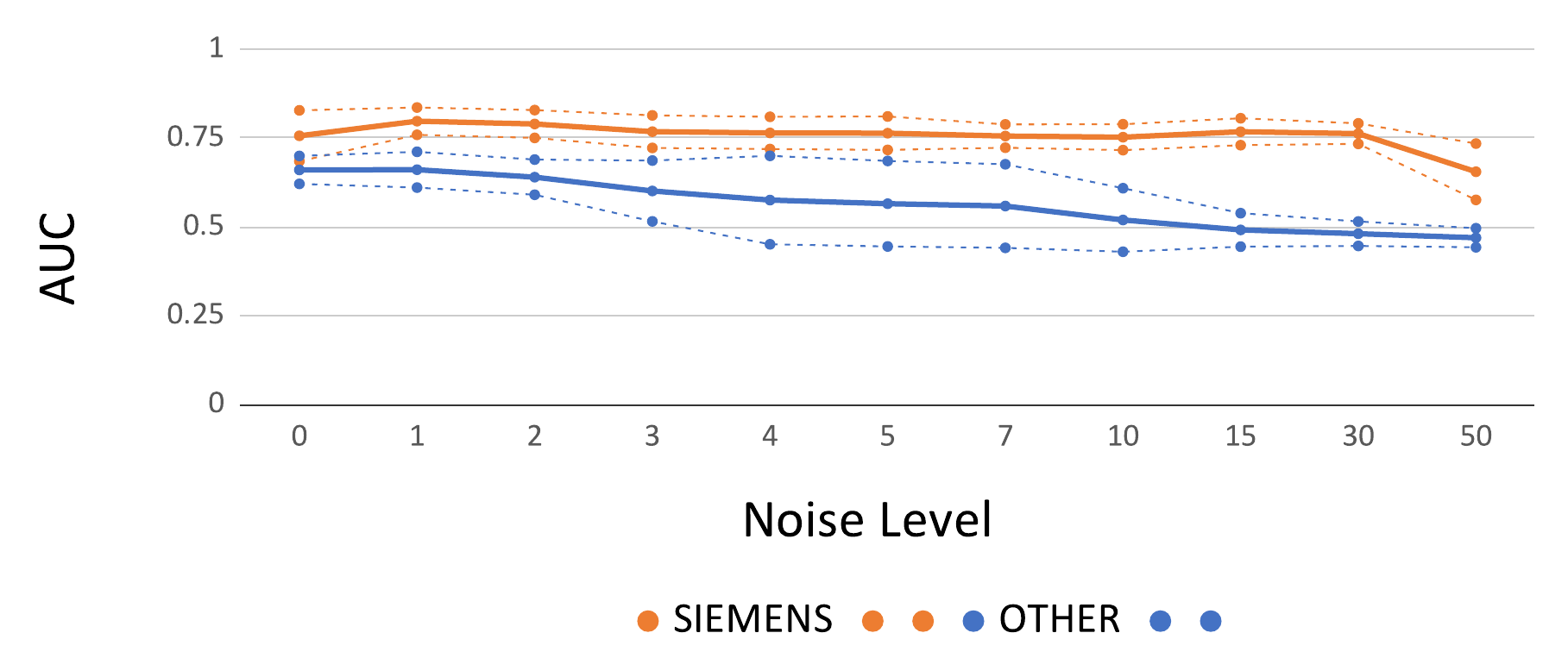}
\includegraphics[width=0.49\linewidth]{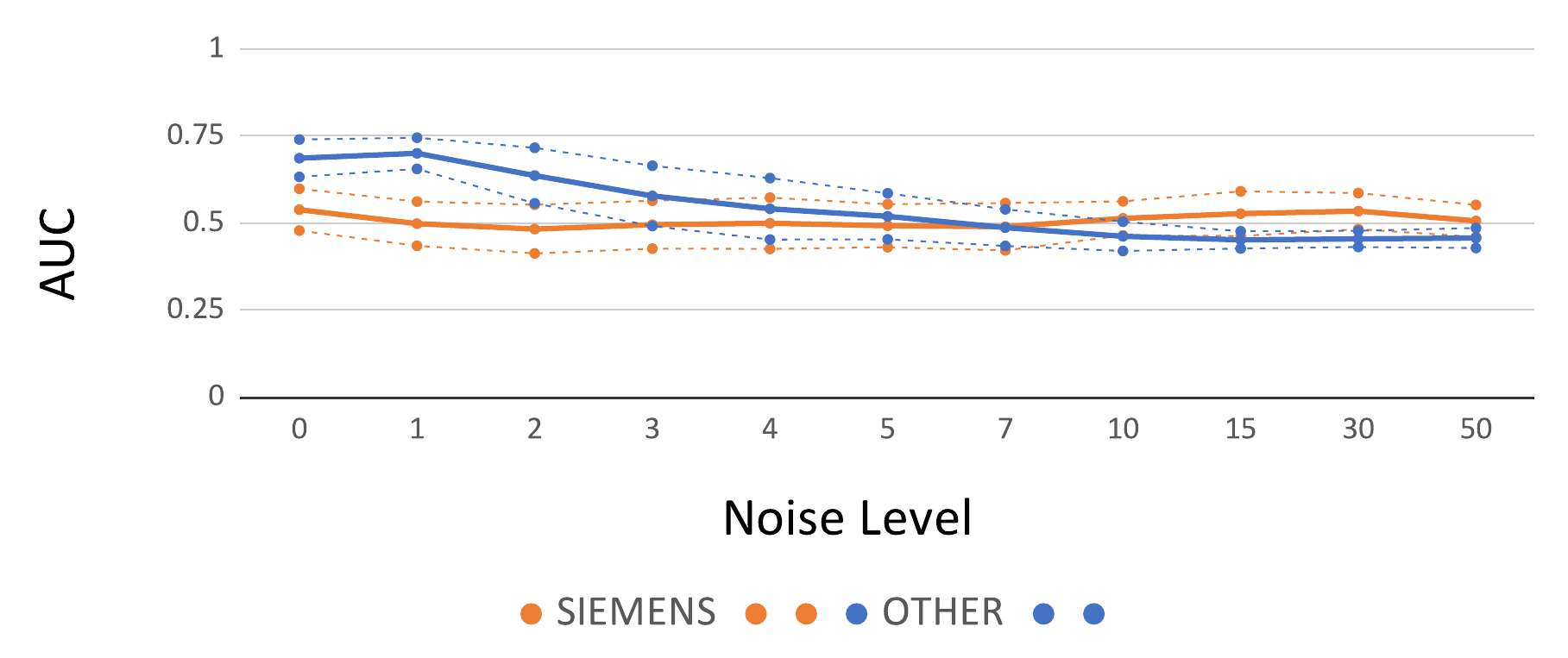}
\caption{Additive noise experiment results for \textbf{(a) CT-Kidney} with Siemens and Other training domains on the left and right, respectively; \textbf{(b) CT-LIDC} with GE and Other training domains on the left and right; and \textbf{(c) MRI-Prostate} with Siemens and Other training domains. Different colors correspond to different scanner domains for the test set. Confidence intervals (standard deviation over all runs) shown with dotted lines.}
\label{fig:noise}
\end{figure*}

We show all results for the noise experiments in Figure \ref{fig:noise}. In fact, adding noise to training images never assisted out-of-domain generalization, even at small noise levels, which, in theory, could have added robustness to the network's learned representations. Instead, adding noise hurt both in-domain and out-of-domain performance, as the noised images no longer accurately represent samples from \textit{either} domain. This shows that more complicated methods are needed to modify images to have domain-generalizable features.

\end{document}